\documentclass[11pt]{article}
\usepackage{epsfig}

\newcommand{\be}{\begin{equation}}
\newcommand{\ee}{\end{equation}}
\newcommand{\ba}{\begin{array}}
\newcommand{\ea}{\end{array}}
\newcommand{\bc}{\begin{center}}
\newcommand{\ec}{\end{center}}
\newcommand{\bi}{\begin{itemize}}
\newcommand{\ei}{\end{itemize}}

\newcommand{\disregard}[1]{{}}

\def\bild#1\over#2{\mathrel{\mathop{\kern0pt #1}\limits_{#2}}}

\begin{document}
\ 
\vskip 2cm
\centerline{\large {\bf LIFSHITZ-LIKE ARGUMENT FOR LOW-LYING STATES}}
\centerline{\large{\bf IN A STRONG MAGNETIC FIELD}}
\vskip 2cm                 
\centerline{\bf Cyril FURTLEHNER\rm } 
\hfill\break
\centerline{\it Institute of Physics, University of Oslo}
\centerline{\it P.O. Box 1048 Blindern}
\centerline{\it  N-0316 Oslo, Norway}

\vskip 2cm
{\bf Abstract} :

 We are interested in the question of the localization of an electron moving in two dimensions,
 submitted to a strong magnetic field and scattered by randomly
 distributed zero-range impurities.
 Considering the  explicit expression for the density of states obtained by Br\'ezin, Gross and Itzykson, we adapt the Lifshitz argument,
 in order to analyse the somewhat unusual power-law behavior of the low energy spectrum. The  typical configurations of disorder
 which gives rise to low energy states are identified as cluster of impurities of well defined form, when the 
 impurity density is smaller than the Landau degeneracy. 
 This allows for an interpretation of
 low lying states, localized around these clusters. The size of these clusters diverges logarithmically when the energy goes to zero.

\vfill\eject

\hfill\break
{\bf 1. Introduction}\\
\hfill\break
 
 The two dimensionnal problem of an electron submitted to a strong magnetic field and moving in  a random potential, 
 has been the subject of intensive investigations, because of its relevance for the integer quantum Hall effect.
 In the case of a locally correlated disordered potential, some explicit results have been found, concerning the average 
 density of states 
 \cite{WGR,BGI,KLP}. Although the DOS doesn't contain in general  any information about localization, exception should be made
 for the tails of the spectrum, which are generally associated with unprobable realizations of the random potential, and as in the
 Lifshitz-tail examples, are interpreted in terms of localized states.   
 In the strong magnetic field problem, such a situation is encountered  in the case of gaussian fluctuations, where the spectrum 
 displays a gaussian tail at large energy \cite{WGR}. If disorder is realized by  delta impurities obeying Poisson statistics,
 the situation is very different. The spectrum is bounded from below, and instead of having a tail, it is singular at low energy.
 More precisely, depending on a parameter $f={\rho\over\rho_l}$ which is the ratio between the density of impurities and the Landau
 degeneracy, it takes the following asymptotic form \cite {BGI},
 \be \label {CBE} \lambda \rho(E)\sim_{\omega\to +0}  \cases{(1-f)\delta
 (\omega)+A(f)\omega^{-f},&$0<f<1$\cr
 {1\over\omega(\ln[\omega/\alpha])^2},&$f=1$\cr
 B(f)\omega^{f-2},&$1<f<2$\cr
 constant,&$f=2$\cr
 C(f)\omega^{f-2},&$f>2$\cr} \ee
 with $\omega={f\over\lambda\rho}(E-\omega_c)$.
 This behavior is very uncommon, and seems to be 
 particular to the choice of short-range single impurity potential, for long range one, one recovers the 
 usual Lifshitz tail \cite{BHKL}.
 In the standard Lifshitz
 argument, when there is no magnetic field,
 low energy states  are localized in regions of space where impurities are absent; 
 An empty region, of typical size  $\pi R^2$, contains states with energy of the order of $1/(\pi R^2)$.
 For a Poisson distribution, the probability of not finding a single impurity in a volume 
 $\pi R^2$ is $\exp (-\rho\pi R^2)$. Identifying the energy to the inverse size of the empty region
 let to obtain the low energy behavior $\exp (-{\rho\over E})$.
 This heuristic argument \cite{LFS} has been
 later confirmed by an exact calculus
 \cite{LGRT,LGRW}.
 The question is whether it is possible to adapt this argument 
 for the problem with strong magnetic field, in order to have a physical interpretation of the base
 of the spectrum, known from
 elsewhere to be constituated of localized states \cite{DMP}.

 Let us consider the case where the density of impurities is less than the Landau degeneracy ($f<1$).
 The zero energy delta peak has a simple interpretation \cite{AZB} and corresponds to the delocalized state which is expected at 
 the center of each Landau band \cite{PSK,ARB} :\\
 these states are indeed linear combinations of Landau states, which vanish at the position of the impurities.
 And in a given volume $V$, the number of Landau states at disposal is $\rho_l V$. The number of
 constraints imposed on the zero energy states is $\rho V$, the number of impurities.
 As a consequence, the corresponding subspace of states has the dimension $(\rho_l-\rho)V$ (unless as will be seen later that two
 impurities coincide). This gives as expected the degeneracy  $\rho_l(1-f)$ per unit volume, given by (\ref {CBE}).
 What remains to be analysed is the $\omega^{-f}$ behavior of the excited states spectrum.

 The paper is organized as follows, in the first part, the problem with a finite number $N$ of impurities is analysed in details.
 The zero modes are first extracted from the Hilbert space, which allows
 then to define the restriction of the Hamiltonian to the excited 
 subspace as a $N\times N$ matrix. The two impurity case is explicitely solved and elucidates the  
 mechanism which produces low energy states. The generalization 
 to a cluster of impurities is then considered and an approximate expression of the lowest energy
 is found. In the second part, a statistical analyses is performed, using this expression, in order  to find the most probable
 configurations corresponding to  a given low  energy, and the  contribution
 to the DOS is computed in the case $f<1$.\\
\hfill\break
 {\bf 2. The $N$ delta impurity problem}\\
{\large\it a. Coherent states basis for the excited subspace}\\

 The $N$ impurity problem, projected onto the LLL is defined by the Hamiltonian
\be H=\lambda P_0\sum_{i=1}^N\delta({\bf r}-{\bf r}_i)\ P_0 \ee
after shifting the spectrum by a constant. $\lambda$ is the coupling constant of the delta potential, $P_0$ is the projection operator
on the LLL.
 Let us consider the basis corresponding to the symmetric gauge,
 centered at position $a$ (using complex notation and magnetic units) :
 \be \phi_p^{a} ({\bf r})={1\over\sqrt{\pi p!}} (z-a)^p\
 e^{-{1\over2}(z\bar z+a\bar a-2z\bar a)}
\qquad p\in N\ee
 In the situation where there is only one impurity, situated at position $a$, these states remain eigenstates, with zero energy for
 $p>0$ and with energy ${\lambda\over\pi}$ for $p=0$.
 Let us associate to the impurity $i$ the coherent state $\psi_i$, corresponding to the only non-vanishing state at ${\bf r}_i$,
 \be \label {EII} \psi_i({\bf r})=\phi_0^{z_i}({\bf r})={1\over\sqrt\pi }\
 e^{-{1\over2}(z\bar z+z_i\bar z_i-2z\bar z_i)} \ee
 As already mentionned, the LLL is divided into two orthogonal subspaces :
 the zero energy subspace of dimension higher or equal to $\rho_l V-N$, and the excited subspace of dimension less or equal to $N$.
 The subspace of wave-functions
 vanishing at ${\bf r}_i$, is orthogonal to $\psi_i$ and contains the zero energy states. Therefore the zero-energy subspace
 is orthogonal to the one generated by $\psi_1$,\ldots $\psi_N$. Let us find under which conditions theses states are linearly
 independents.\\
 Consider 
 \be \label {psia} \psi({\bf r})=\sum_{i=1}^N a_i\psi_i({\bf r}) \ee
 a linear combination of these $N$ states. In the Landau symmetric basis  $\psi$ has the form
 \be \label {psib} \psi ({\bf r})=\sum_{p=0}^{\infty} b_p \phi_p({\bf r})\ee
 The relation between the $b_p$ and $a_i$ is
 \be b_p={1\over\sqrt p!}\sum_{i=1}^N a_i \bar z_i^p\
 e^{-{1\over2}z_i\bar z_i} \ee

 In order for  $\psi$ to be identically zero, the $b_p$ have to vanish.
 In particular, imposing this to the first $N$  ($p=0\ldots N-1$) leads to an homogeneous system of equations for the $a_n$,
 with a determinant proportionnal to the $\bar z_i$'s Vandermonde determinant, i.e. a completely antisymmetric function of these 
 variables. Therefore a necessary condition for the  $\psi_i$ to be linearly  dependent is that two impurities coincide, and it
 is evidently sufficient. As a consequence $\psi_1$,\ldots$\psi_N$
 is a basis of the excited subspace (non-orthogonal).\\
 Let us write the Hamiltonian into this basis. Starting from the decomposition (\ref {psia}) of an arbitrary excited state, 
 the action of $H$ on this state
 is
 \be <{\bf r}|H|\psi>=\lambda \sum_{i=1}^N P_0({\bf r},{\bf
 r}_i)\sum_{n=1}^N a_n\psi_n({\bf r}_i)\ee
 with
 \be P_{0}({\bf r},{\bf r}')={1\over\pi }\
 e^{-{1\over2}(z\bar z+z'\bar z'-2z\bar z')} \ee
 the kernel of the LLL projection operator. Using the fact that
 \be <\psi_i|\psi_j>=\pi P_{0}({\bf r}_i,{\bf
 r}_j)=\sqrt\pi\psi_j({\bf r}_i) \ee
 we obtain
 \be <{\bf r}|H|\psi>=\lambda \sum_{i=1}^N\sum_{j=1}^N a_j P_{0}({\bf r}_i,{\bf
 r}_j)\psi_i({\bf r})\ee
 In conclusion the matrix elements of $H$ in the $(\psi_1,\ldots ,\psi_N)$ basis are given by
 $\lambda P_0({\bf r}_i,{\bf r}_j)$.\\
 \hfill\break
{\large\it b. Two impurities}\\

 In the case with only two impurities, we can diagonalize this matrix. Choosing the spatial reference such that 
 $z_1=-z_2=a/2$, where $a$ is the distance between the two impurities, 
 we have
 \be H_2={\lambda\over\pi }\pmatrix{1&e^{-{1\over2}a^2}\cr
 e^{-{1\over2}a^2}&1\cr}\ee
 The eigenvalues of this matrix correspond to the energies $E_-$ and  $E_+$ of the two excited states, 
 \be E_{\pm}={\lambda\over\pi }(1\pm\ e^{-{1\over2}a^2}) \ee
 The corresponding wave-functions beeing (up to a normalization coefficient)
 \be \psi_{\pm}=\psi_1 \pm \psi_2
 ={1\over\sqrt\pi}e^{-{1\over2}(z\bar z +a^2)}(e^{2az}\pm e^{-2az})\ee
 In conclusion, when the two impurities are well seperated, the excited states have almost the same energy, comparable to the one
 impurity value. Whereas, a low energy state is obtained when the two impurities are close.
 For $a<<1$ this energy behave like
 \be\label{E2} E_-\simeq {\lambda\over2\pi }a^2 \ee
 \hfill\break
{\large\it c. Impurity cluster}\\

 The preceeding example suggests that low-energy states are associated with regions of high concentrations of impurities.
 Indeed, $N$ impurities involve $N$ localized states $\psi_i$ (which have a characteristic size $1/\rho_l$).
 Low-lying states are expected to appear when the overlap between these states starts to be important.
 Consider a situation (figure 4) where $N$ impurities are located in a small volume ($\pi R^2$), that is
 a cluster of impurities, then the 
 $N$ corresponding states overlap essentially with the $N_l=\rho_l\pi R^2$ Landau states situated inside the disc (in the symmetric
 gauge, the states of the LLL are localized on a ring of radius $\sqrt{l/\rho_l\pi}$,
 where $l$ is the angular momentum \cite{ITZD})
 So if $N>N_l$, we expect to have $N-N_l$ low-energy states. 
 It seems therefore natural to consider such configurations in order to analyse the bottom of the spectrum.\\
 Let us estimate the lowest energy corresponding to such a configuration.
 Consider the decompositions (\ref {psia}) and  (\ref {psib}) of an excited state. A low energy state is supposed
 to avoid the impurities. A way to construct such a state, is to impose on the  $b_p$ to vanish until $p=N-2$ included.
 In that case, $\psi$ has components only on the Landau states  $p>N-2$, situated at a distance from the center of the cluster
 greater or equal to $\sqrt{N-1\over\pi\rho_l}$.
 Such a state is given by the $a_n$, solution of the set of equations
 \be b_p=0=\sum_{n=1}^N a_n e^{-{1\over2}z_n\bar z_n}\ z_n^p \qquad p=0,\ldots
 N-2 \ee
 And the solution, up to a proportionality constant is 
 \be a_n e^{-{1\over2}z_n\bar z_n}=C_{N,n} \ee
 where $C_{N,n}$ is the cofactor of the element $(N,n)$ in the 
 Vandermonde type matrix:
 \be D_N^p=\pmatrix{1&\ldots&\ldots &1\cr
                    z_1&\ldots&\ldots &z_N\cr
		    \vdots&\ldots&\ldots&\vdots \cr
		    z_1^{N-2}&\ldots&\ldots &z_N^{N-2}\cr
		    z_1^p&\ldots&\ldots &z_N^p\cr}
		    \ee
 In particular, for $p=0\ldots N-2$,
 \be \det\ D_N^p=0=\sum_{n=1}^N C_{N,n}\ z_n^p\ee
 which is precisely what we want.
 Moreover the $C_{N,n}$ have the expression
 \be C_{N,n}=(-1)^{{N(N-1)\over2}+n}\prod_{p<q\ p,q\ne n} (z_p-z_q) \ee
%\hfill\break
% \input figure24.pstex_t
% \hfill\break
% \hfill\break
 The matrix $H_N=\lambda P_0({\bf r}_i,{\bf r}_j)$, written in 
 $\psi_1,\ldots \psi_N$ basis, is self-adjoint and positive, so its smallest eigenvalue $E_0$ verify the inequality:
 \be E_0\le {(\psi|H_N\ \psi)\over (\psi|\psi)} \ee
 with the norm defined by,
 \be (\psi|\psi)=\sum_{n=1}^N \bar a_n a_n  \ee
 From this choice and for the considered state the inequality rewrites
 \be E_0\le E={\lambda\over\pi}
 \ {\sum_{n,m}\ \bar C_{N,n}\ C_{N,m} e^{z_m\bar z_n}\over \sum_n
 \ |C_{N,n}|^2\ e^{|z_n|^2}}
 \ee
 and 
 \be E\le{\lambda\over\pi}\  {\sum_{n,m}\ \bar C_{N,n}\ C_{N,m} e^{z_m\bar z_n}\over \sum_n
 \ |C_{N,n}|^2\ }
 \ee
 Expanding the exponential in the preceeding expression, we observe that the first non-vanishing term
 corresponds to ${(z_m\bar z_n)^{N-1}\over(N-1)!}$, because the determinant of $D_N^p$ is zero for $p<N-1$.
 In addition, since  $|z_n|^2\le N_l\le N$, the serie has a rapid decay, which allow to neglect the remainder of the expansion.
 We then obtain
 \be E_0\le{\lambda\over\pi}\ {1\over (N-1)!}{|D_{N-1}|^2\over\sum_{n=1}^N |C_{N,n}|^2}\ee
 with  $D_{N-1}=(-1)^{N(N-1)\over2}\prod_{p<q}(z_p-z_q)$ the
 Vandermonde determinant of the $z_n$ variables.
 If $n^{*}$ labels the impurity for which, 
 $\prod_{p\ne n}|z_n-z_p|^2$ is minimum, then we have the inequality
 \be \sum_{n=1}^N |C_{N,n}|^2\ge N|C_{N,n^*}|^2 \ee
 which leads to the approximate form for $E_0$\\
% Soit $D_p$ le d\'eterminant de la matrice $D_N^p$. La fonction
% g\'en\'eratrice de ces d\'eterminants est donn\'ee par :
% \be \sum_{p=0}^{\infty}\alpha^p D_p=\alpha^{N-1} {D_{N-1}\over\prod_{n=1}^N
% (1-\alpha z_n)} \ee

 \be \label {FAL} E_0\propto{\lambda\over\pi}
 \ {1\over N!}\min_{p}\prod_{n\ne p}|z_p-z_n|^2 \ee
 and which coincides with expression (\ref{E2}) in  the two impurities case.\\
\hfill\break
\hfill\break 
 {\bf 3. Cluster thermodynamic}\\ 
 
 We can now use this expression in order we understand the low energy behavior
 of the spectrum obtained by Br\'ezin, Gross,
 Itzykson
 ($f<1$).
 We start from the principle that each impurity in the system gives rise to an excited state
 with energy depending on the configuration of the other impurities.
 If the concentration around one impurity is high, in other words if the impurity is in a cluster,
 then the corresponding energy is low and not affected by the impurities situated outside of the cluster (too far away 
 for overlap effect).
 We can therefore associate a low-energy state to the formation of a cluster around an impurity, and by extension,
 a density of states per impurity.
 Let $X_i$ be a variable parametrizing the cluster configuration of the impurity $i$.
 Its contribution to the density of state per impurity is proportionnal to the probability  $P(X_i)$
 of being realized
 \be \label {DOS} \rho_i(E)=\int DX_i\ P(X_i)\ \delta(E(X_i)-E) \ee
 So in average, the low-energy density of states by unit volume is proportional to  $\rho$ times the preceeding expression.
 If we use now the expression (\ref {FAL}) to evaluate the energy of the clusters, we see that to a given energy
 corresponds a statistical ensemble of clusters. 
 Each cluster is defined by its volume $N_l$,
 its mean density $\nu=N/N_l\ >1>f$, and by the positions $z_i,\ i=1\ldots N$, of the impurities
 in the cluster.
 At very low energy, the clusters are expected to be macroscopic objects, and have to be described by a finite number of macroscopic
 variables, giving the density profile, in replacement of the microscopic degrees of freedom (namely the
 individual positions of impurities). 
 Let us look first for the distribution of positions in a cluster of energy $E$, size $N_l$ and mean density $\nu$.
 For a given configuration the energy is
 \be E=e^{\sum_{n=1}^N\ \log|z_n|^2\ -N\log N +N} \ee
 using the Stirling formula ($N!\simeq N^N\ e^{-N}$) and
 with $0\le |z_n|^2\le N_l$ ($N_l=\pi\rho_l R^2$).
 Consider a subdivision of the cluster in  $M$ cells, corresponding to intervals of the $|z_n|^2$ equal to $a=N_l/M$
 (cells with identical area ${\pi R^2\over M}={a\over\rho_l}=\pi\delta r^2={\delta
 |z|^2\over\rho_l}$).
 If $n_p$ is the number of impurities in the cell $p$, then the probability associated to this configuration $(n_1,\ldots ,n_M)$ 
 of the cluster
 is
 \be P(n_1,\ldots ,n_M,N)={N!\over n_1!\ldots n_M!}({1\over M})^N\
 {(fN_l)^N\over N!} e^{-fN_l} \ee
 Since we are interested in the continuum limit,  define ($x=|z|^2/N_l={p\over N_l}a$)
 \be \nu(x)dx=n_p=\nu(x){1\over M}\qquad 1\ll M\ll N \ee
 the energy takes then the form
 \be \log E=N_l\int_0^1[\nu(x)\log x +\nu-\nu\log\nu]dx\ee
 and, at leading contribution in $N_l$, the probability is
 \be \log P=N_l\int_0^1 [\nu(x)(1-\log{\nu(x)\over f})-f] dx\ee
 with the constraint
 \be \int_0^1 \nu(x)dx=\nu \ee
 Let us determine the configuration for which  $\log P$ is maximum at fixed $E$, $N_l$ and $\nu$. Using a Lagrange multiplier for 
 the energy constraint we obtain the saddle point equation
 \be {\partial\log P\over \partial\nu(x)}-\alpha {\partial\log E\over \partial
 \nu(x)}=0\ee
 The solution, with proper normalization, is
 \be \nu(x)=\nu (1-\alpha)({x\over N_l})^{-\alpha} \ee
 $\alpha$ being implicitely determined through the relation between $\gamma={1\over 1-\alpha}$
 and the energy,
 \be \label {EG} \log E=-N_l(\nu(\log\nu-1) +\gamma\nu) \ee
 and the probability now takes the form 
 \be \log P=\log E\ -N_l(f-\nu(1+\log\gamma f))\ee
 At a given energy, the possible configurations are parametrized by $(N_l,\nu)$. The saddle point is determined by the 
 set of equations  
 \be \Big ({\partial\log P\over \partial N_l}\Big)_{\nu,E}=0 \ee
 \be \Big ({\partial\log P\over \partial \nu}\Big)_{N_l,E}=0 \ee
 Using (\ref{EG}), which determines $\gamma$ these equations rewrites
 \be \log\gamma f\ +{1-\log\nu\over\gamma}-{f\over\nu}=0 \ee
 \be \log\gamma f\ -{1\over\gamma}\log\nu=0 \ee
 So finally, $\log P$ has its maximum (which can be verified by computing second derivatives) at energy $E$ when 
 $\nu=1$,
 $N_l=-f\log E/(1-f)$, which corresponds to $\gamma=1/f$. The other solution ($\nu=f$ and $\gamma=1$ is also a local maximum, but
 outside the range of interest for the parameters. 
 For this type of configurations (paramatrized now only by $X=N_l$), we have the relation
 \be \log{P\over E}=-f\log E \ee
 Using (\ref {DOS}) (with the change of variable $DX\propto dE/E$),
 we arrive at the expected low-energy behavior of the density of states 
 \be \rho (E)\propto E^{-f} \ee
 Moreover, states contributing to this behavior are associated to the existence of impurity clusters of size
 $N_l=-f\log E/(1-f)$ and the shape
 \be \nu(x)= f\  x^{f-1} \ee
 In contrary to the Lifshitz argument,  the low energy states are associated 
 with regions of high impurity concentration around which they localize, with caracteristic size $\log{1\over E}$,
 a rough indication of a logarithmic divergency of the localization length, at least when $f\ll 1$.
 This feature  might be very particular to the zero-range nature of the impurity scattering potential.
 When $f$ approaches $1$, this picture might be modified by some ``percolation'' effect of these cluster.
 For $f$ greater than $1$ the argument developped in this paper is not applicable, neither is the stantard  Lifshits argument,
 to reproduce the low energy spectrum.
 This seems to indicate that states are not localized at the bottom of the spectrum in this case.

\hfill\break
\hfill\break 
 {\bf Acknowledgements}\\ 
 I am gratful to A. Comtet for interesting me to this question and for useful discussions, as well as 
 to J.M. Leinaas, L. Pastur and  S. Villain-Guillot.
 I am thankful   to  the Institute of physics of the Oslo University for hospitality, and 
 the National Reasearch Council of Norway for financial support.

  {}

\end{document}